\documentstyle[psfig]{mn}
\def\lsim{\mathrel{\hbox{\rlap{\hbox{\lower4pt\hbox{$\sim$}}}\hbox{$<$}}}}
\def\gsim{\mathrel{\hbox{\rlap{\hbox{\lower4pt\hbox{$\sim$}}}\hbox{$>$}}}}
\def\simlt{\mathrel{\rlap{\lower 3pt\hbox{$\sim$}}
        \raise 2.0pt\hbox{$<$}}}
\def\simgt{\mathrel{\rlap{\lower 3pt\hbox{$\sim$}}
        \raise 2.0pt\hbox{$>$}}}
\newcommand{\kms}{\, {\rm km\, s}^{-1}}

\begin{document}  
        
\title[The distribution of supermassive black holes in the 
nuclei of nearby galaxies]
{The distribution of supermassive black holes in the nuclei 
of nearby galaxies}
\author[Andrea Cattaneo, Martin G. Haehnelt and Martin J.  Rees]
{Andrea~Cattaneo $^1$, Martin G.~Haehnelt $^{1,2}$ and Martin J.~Rees $^1$\\
$^1$Institute of Astronomy, Madingley Road, Cambridge CB3 0HA\\
$^2$Max-Planck-Institut f\"ur Astrophysik, Karl-Schwarzschild-Str. 1, 
85740 Garching}
     
\maketitle

\begin{abstract}
The growth of supermassive black holes by merging and accretion in 
hierarchical models of galaxy formation is studied by means of Monte 
Carlo simulations. 
A tight linear relation 
between masses of black holes and masses of bulges arises
if the mass accreted by supermassive black holes scales linearly 
with the mass forming stars and if the redshift evolution of mass
accretion tracks closely that of star formation. 
Differences in redshift evolution between black hole accretion and star 
formation introduce a considerable scatter in this relation.  
A non-linear relation between black hole accretion and star formation 
results in a non-linear  relation between  masses of remnant black
holes and masses of bulges.
The relation of  black hole mass to bulge luminosity observed
in nearby galaxies and its scatter
are reproduced reasonably well
by models in which black hole accretion and star formation are linearly   
related but do not track each other in redshift.
This suggests that a common mechanism determines
the efficiency for black hole accretion and the efficiency for 
star formation, especially for bright bulges.    
\end{abstract}


\section{Introduction}  

Lynden-Bell (1969) suggested that active galaxies are fuelled by
accretion of matter onto supermassive black holes and that 
massive dark compact objects could therefore be present in the 
nuclei of many quiescent galaxies as remnants of QSOs.  
Now there is solid 
observational evidence for the presence of supermassive 
black holes in our own and in many nearby galaxies.
A correlation  between the mass of a supermassive black hole and the 
luminosity of the bulge of the host galaxy was already apparent in the
first small sample of detections. The larger samples meanwhile 
available have backed up the existence of such a correlation
(Kormendy 1993; Kormendy \& Richstone 1995; Magorrian et al. 1998; 
Ford et al. 1998; van der Marel 1998; Ho 1998; Mc Leod 1998). 

The growth  history  of these supermassive black holes is, however,  rather  
uncertain. It is widely believed that they built up during 
the QSO epoch, around $z\sim 2-3$. As pointed 
out by Phinney (1997) and discussed in more detail by Haehnelt, Natarajan \& 
Rees (1998, HNR98), the total integrated mass density in remnant black holes 
substantially exceeds the integrated mass density accreted onto supermassive 
black holes by optically bright QSOs unless either the efficiency for producing
radiation is rather small or the masses of black holes in nearby galaxies are 
overestimated (see also Richstone et al. 1998 for a review).  
This leaves some room for accretion other than that inferred 
from optically bright QSOs (see also Fabian \& Iwasawa 1999 
and Salucci et al. 1999).  

A striking feature of the activity of optically bright QSOs is the fast 
evolution of their space density. A rapid rise between $z\sim5$ 
and $z\sim3$  is followed by a strong decline after $z\sim 2$ 
(e.g. Shaver et al. 1996). This rapid rise can be plausibly 
explained by the evolution of the dark matter (DM) distribution in
hierarchical cosmogonies (Haehnelt \& Rees 1993, HR93):
deeper potential wells form at later times 
and  the first potential wells which are deep enough to retain 
their gas in spite of the input of energy and momentum due to star formation 
offer the best conditions for efficient formation of supermassive
black holes. As discussed by HR93 and Cavaliere, Perri \& Vittorini (1997), 
the decline in the activity of galactic nuclei is probably due to a transition 
from a high-redshift phase, in which supermassive black holes form in 
initially gas rich nuclei, to a low-redshift phase, where QSOs are 
refuelled by mergers of increasingly gas-poor galaxies. 
The late merging of gas-poor galaxies containing supermassive black holes will 
obviously affect the mass distribution of remnant black holes. For instance, it
will introduce scatter in any relation between properties of galaxies 
and masses of remnant black holes. Here we use Monte Carlo realizations
of the merging histories of DM haloes, together with a simple scheme 
for galaxy formation, to model the merging of galaxies in a 
hierarchical cosmogony. We investigate different schemes 
for the accretion of matter onto supermassive black holes and  
we study the effect on the mass distribution of remnant black holes
in low-redshift galaxies. 
The structure of the paper is as follows.
In Section 2 we describe the Monte Carlo simulations. 
In Section 3 we compare the resulting relations between masses of black holes 
and luminosities of bulges to that observed in nearby galaxies. 
Section 4 contains our conclusions. We assume a flat cosmology
with $\Omega_M=0.3$, $\Omega_\Lambda=0.7$ and a Hubble constant 
of $75{\rm\,km\,s^{-1}\,Mpc^{-1}}$.

\section{Monte Carlo simulations of the merging and accretion history
of supermassive black holes}

\subsection{Merging histories of dark matter haloes}

The standard paradigm for the formation of structure in the Universe
is the gravitational instability of density fluctuations in a primordial
Gaussian random field. In this picture, the number density of collapsed 
DM haloes is accurately described by the Press-Schechter formula 
\begin{eqnarray}
&N(M,z){\rm\,d}M=\\ \nonumber
&-\left({\overline{\rho}\over M}\right)
({1\over 2\pi})^{1\over 2}\left({\omega\over\sigma}\right)
\left({1\over\sigma}{{\rm d}\sigma\over{\rm d}M}\right)
{\rm exp}\left(-{\omega^2\over 2\sigma^2}\right){\rm\,d}M\\ \nonumber
\end{eqnarray}
(Press \& Schechter 1974).
Here $\overline{\rho}(z)$ is the cosmological density at redshift $z$, 
$\sigma_0(M)^2$ is the variance of the linearly extrapolated density field 
on the scale $M$, 
$\omega$ is defined as $\omega(z)\equiv\delta_{\rm c} D(0)/D(z)$,
where $\delta_{\rm c}$ 
is the over-density threshold at which density fluctuations collapse,
and $D(z)$ is the linear growth factor of density fluctuations.  
The variance of the density field is related to the power spectrum of the 
density fluctuations. 
Here we assume a standard cold dark matter power spectrum as given by 
Bond \& Efstathiou (1984), with a normalisation that reproduces  the
present-day space density of clusters of galaxies (Eke, Cole \& Frenk 1996). 
The dependence of the linear growth factor on redshift is given by 
Heath (1977).

Techniques for generating Monte Carlo realizations of the merging
histories of DM haloes based on extensions of the Press-Schechter 
formula have been described for instance
by Cole \& Kaiser (1988), Cole (1991),
Lacey \& Cole (1993), Kauffmann \& White (1993) and Somerville \&
Kolatt (1997). The basic  
equation is the conditional probability 
that a halo of mass $M$ at redshift $z$ has a progenitor of 
mass $M-{\Delta M}$ at redshift $z+\Delta z$,
\begin{eqnarray}
&P(M\rightarrow M-\Delta M,z\rightarrow z+\Delta z)=\\ \nonumber
&\frac{1}{\sqrt{2\pi}}\;
{\omega(z+\Delta z)-\omega(z)\over[\sigma_0^2(M-\Delta
M)-\sigma_0^2(M)]^
{3\over 2}}\; \exp{\biggl  \{ -{[(\omega(z+\Delta z)-\omega(z)]^2\over
2[\sigma_0^2(M-\Delta M)-\sigma_0^2(M)]} \biggr \}} \\\nonumber
\end{eqnarray}
(Lacey \& Cole 1993).  We follow the  description of 
Somerville \& Kolatt (1997) to construct Monte Carlo realizations
of the merging histories of DM haloes from equation (2). 
The probability distribution (2) is used to assign a mass $M$ to 
a progenitor at redshift $\Delta z$ of a DM halo 
of mass $M_0$ at $z=0$.  The  procedure is 
iterated and further progenitors are drawn from the distribution (2), 
but the merging history  is followed only for haloes with circular 
velocities $v_{\rm c}=({\rm G}M/r_{\rm vir})^{1/2}$
above a certain threshold $v_l$. 
Haloes with $v_{\rm c} <v_{\rm l}$ are treated as accreted mass.
Progenitors with a mass larger than the not yet allocated mass are discarded. 
We construct progenitors of progenitors until all haloes have 
$v_{\rm c} <v_{\rm l}$.   
The procedure is repeated for a representative sample of haloes at $z=0$.   
We adopt a step of $\Delta z=0.01$ and a resolution of 
$v_{\rm l}=70{\rm\,km\,s}^{-1}$.

\subsection{The galaxy formation scheme}

The modelling of galaxies within hierarchal cosmogonies has reached 
a considerable level of complexity. Here we concentrate on the 
effects of mergers on the growth of supermassive black holes 
and on the distribution of remnant black holes.  Therefore we adopt a
rather simple scheme  for galaxy formation similar to that proposed
by  White and Rees (1978), which should nevertheless catch the
essential features of the hierarchical merging of galaxies (White
1996). 

Important conditions for the 
formation of a galaxy in a collapsed DM halo are
the ability of the gas to cool and the ability of the halo 
to retain its gas in spite of the input of energy and momentum
due to star formation and supernova explosions. 
The ability of the gas to cool depends on the temperature and  the density of 
the gas. For cooling by Bremsstrahlung these  dependencies conspire to give
an upper limit for the mass of the gas which is able to cool efficiently 
(Silk 1977; Rees \& Ostriker  1977). 
Feedback from star formation is more important in haloes with shallower
potential wells and therefore smaller circular velocities
(see e.g. Kauffmann, White \& Guiderdoni 1993, KWG93). 
Here we do not treat the cooling and feedback explicitly. Instead, we 
introduce an effective efficiency for star formation, which depends  
on the halo circular velocity  in such a way that the mass of the galaxy
associated with a 
DM halo of mass $M_{\rm halo}$ at redshift $z$ scales as 
\begin{equation} 
M_\star(M_{\rm halo})= \epsilon_{\star}M_{\rm halo}  v_{\rm c}^{2}
\exp{[-(v_{\rm c}/v_{\rm max})^4]}
\end{equation}
independently of redshift; 
$\epsilon_\star$ and $v_{\rm max}$ are chosen in such a way
that the luminosity function of our simulated 
galaxies reproduces the observed luminosity function of galaxies
at redshift zero (Fig.1).
The cut-off in virial velocities mimics the  inability 
of the gas to cool and form stars in the very deep potential wells
which form at late times.

When DM haloes merge, it will take some time before the  corresponding
galaxies sink to the centre of the merged halo due to dynamical
friction.  To model this, 
we follow KWG93  and assume that initially a single galaxy forms at
the centre of each halo. When haloes merge, the central 
galaxy of the largest progenitor halo becomes the central galaxy
of the new halo. The ``satellite''  galaxies are assumed to  
merge with the central galaxy on the dynamical friction time-scale 
\begin{equation}
t_{\rm df}={1.17r_{\rm vir}^2v_{\rm c}\over{\rm ln}\left({M/M_{\rm sat}}\right)
{\rm G}M_{\rm sat}}
\end{equation}
(Binney \& Tremaine 1987).
Here $M$, $r_{\rm vir}$, $v_{\rm c}$ are the mass, the virial radius and 
the circular velocity of the new  halo and  $M_{\rm sat}$ is the total mass of
the satellite including its DM halo.
The stellar mass of the merged galaxy is assumed to 
be the maximum of the sum of the stellar masses 
of the merging galaxies and the mass given by equation (3). 
If the latter is larger, we interpret the difference  
$\Delta M_{\star}$ as the amount of
gas that has formed stars in the disks of the galaxies
since the last merger. 
  
We follow KWG93 again and introduce a critical mass ratio $f_{\rm e}$
above which a merger produces an elliptical galaxy. 
For mass ratios below the threshold the smaller galaxy is 
assumed to be tidally disrupted and its mass is added
to the disk of the larger one. The value of $f_{\rm e}$ is determined 
by requiring the Monte Carlo simulations to reproduce the 
observed bulge luminosity function, 
once the other free parameters have been fixed. 
A mass-to-light ratio of 
$M_\star/L_B=5(L_B/5.8\times 10^{10}L_\odot)^{0.25}$ is used
(Salucci et al. 1999). Figure 1 compares the resulting 
luminosity functions
to those observed. 

\begin{figure}
\centerline{\psfig{figure=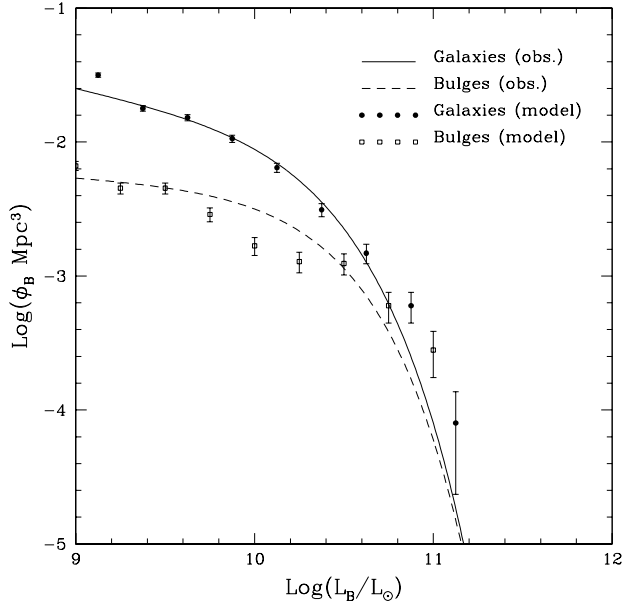,width=0.48\textwidth,angle=0}}
\caption{Observed and simulated  B-band luminosity function 
of galaxies and bulges (data  from Salucci et al. 1998
and Marzke et al. 1998). A relation of stellar mass 
to  circular velocity and mass of the DM halo of the form 
$M_\star(M,z)= 0.014 M_{\rm halo} \times (v_{\rm c}/200\kms) ^{2} 
\exp{[-(v_{\rm c}/300\kms)^4]}$ is assumed. 
Major mergers result in the formation of a bulge 
for mass ratios above $f_{\rm e} = 0.2$.}
\end{figure}

\begin{figure}
\centerline{\psfig{figure=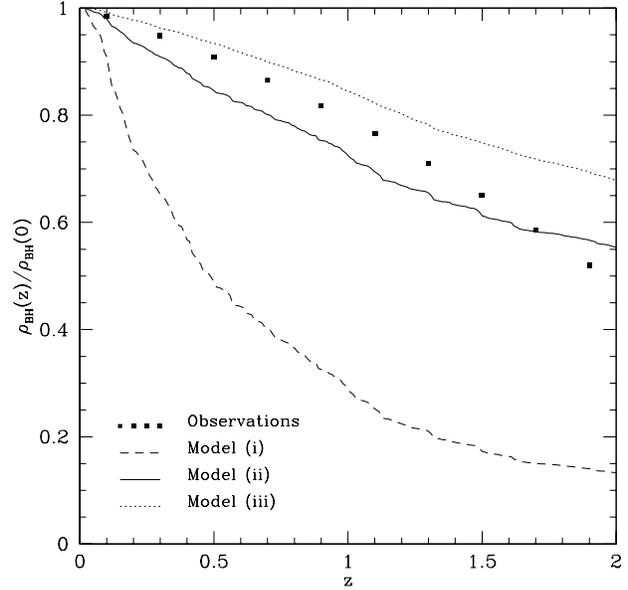,width=0.48\textwidth,angle=0}}
\caption{The redshift dependence of the comoving mass density in 
supermassive black holes for: 
(i) $M_{\rm acc} = 6\times 10^{-3}\Delta M_{\star}$, ({\it dashed line});  
(ii) $M_{\rm acc }= 1.4\times 10^{-3} (1+z)^{2} \Delta M_{\star}$
     ({\it solid line});  
(iii) $M_{\rm acc} = 10^{-6} (1+z)^{2}   M_{\rm halo}
     \exp{[-(v_{\rm c}/300{\rm\,km\,s}^{-1})^4]}$,
     ({\it dotted line}).  
The points show the observed redshift dependence as determined
by Choski \& Turner (1992).}
\end{figure}
      
\subsection{The growth of supermassive black holes by accretion and
merging} 

In a hierarchical cosmogony supermassive black holes grow by merging and by
accretion of gas. Here we assume that 
a major galactic merger leads to an immediate coalescence of the 
supermassive black holes contained in the merging galaxies.  
This should be a reasonable assumption for violent mergers, 
though in a smooth stellar background it may take longer for the holes to 
coalesce (Begelman, Blandford \& Rees 1980). 
Some fraction of the gas which is able to cool in DM haloes is assumed
to be accreted onto the merged supermassive black hole.  
The fraction of cool gas ending up in the hole should
have a complicated dependence on halo and galactic properties
(redshift, past accretion history, past star formation history, 
past merging history, angular momentum). 
We shall not attempt to model these processes here in detail.
We shall rather investigate three very simple models for the 
efficiency of accretion.

\begin{itemize}
\item[(i)]{During a major merger a fixed fraction 
    of the gas that has formed stars since the last merger 
    in the disks of the colliding galaxies is accreted onto the 
    supermassive black hole formed by the coalescence of the 
    supermassive black holes in the merging galaxies
    ($M_{\rm acc} = \epsilon_{\rm acc} \Delta M_{\star}$,
      $\epsilon_{\rm acc}= 6\times 10^{-3}$).}

\item[(ii)]{As in model (i) but with
     an accretion efficiency dependent on redshift  
     to approximate the redshift evolution of the 
     accretion onto  black holes
     inferred from the counts of optically bright QSOs 
    ($M_{\rm acc} = \epsilon_{\rm acc} (z) \Delta M_{\star}$,
     $\epsilon_{\rm acc}(z) = 1.4\times 10^{-3} (1+z)^{2}$).}  

\item[(iii)]{During a major merger a fraction of the total baryonic mass 
      of the merged DM halo is accreted onto the supermassive black hole 
      with an accretion efficiency dependent on redshift as in model (ii)
      and with the same cut-off for high circular velocity
      as in equation (3) 
      to mimic the inability of the gas to cool in deep potential wells
      which form at late times 
      ($M_{\rm acc} = \epsilon_{\rm acc} (z) M_{\rm
      halo}  \exp{[-(v_{\rm c}/300\kms)^4]}$,
      $\epsilon_{\rm acc}(z) = 10^{-6} (1+z)^{2}$).} 
\end{itemize}

Figure 2 shows
the fraction of the  total mass accreted before redshift $z$ 
in the Monte Carlo simulations and
the observed fraction inferred from the blue light 
emitted by optically bright QSOs (Soltan 1982; Choski \& Turner
1992).  Models  (ii) and (iii) approximate the observed 
redshift evolution by construction,
whereas in model (i) most of the mass is accreted at lower redshifts 
than those indicated by the epoch of optically bright QSOs. 
Model (iii) is similar to that recently proposed by Haiman \& Loeb (1998).

\begin{figure*}
\centerline{\psfig{figure=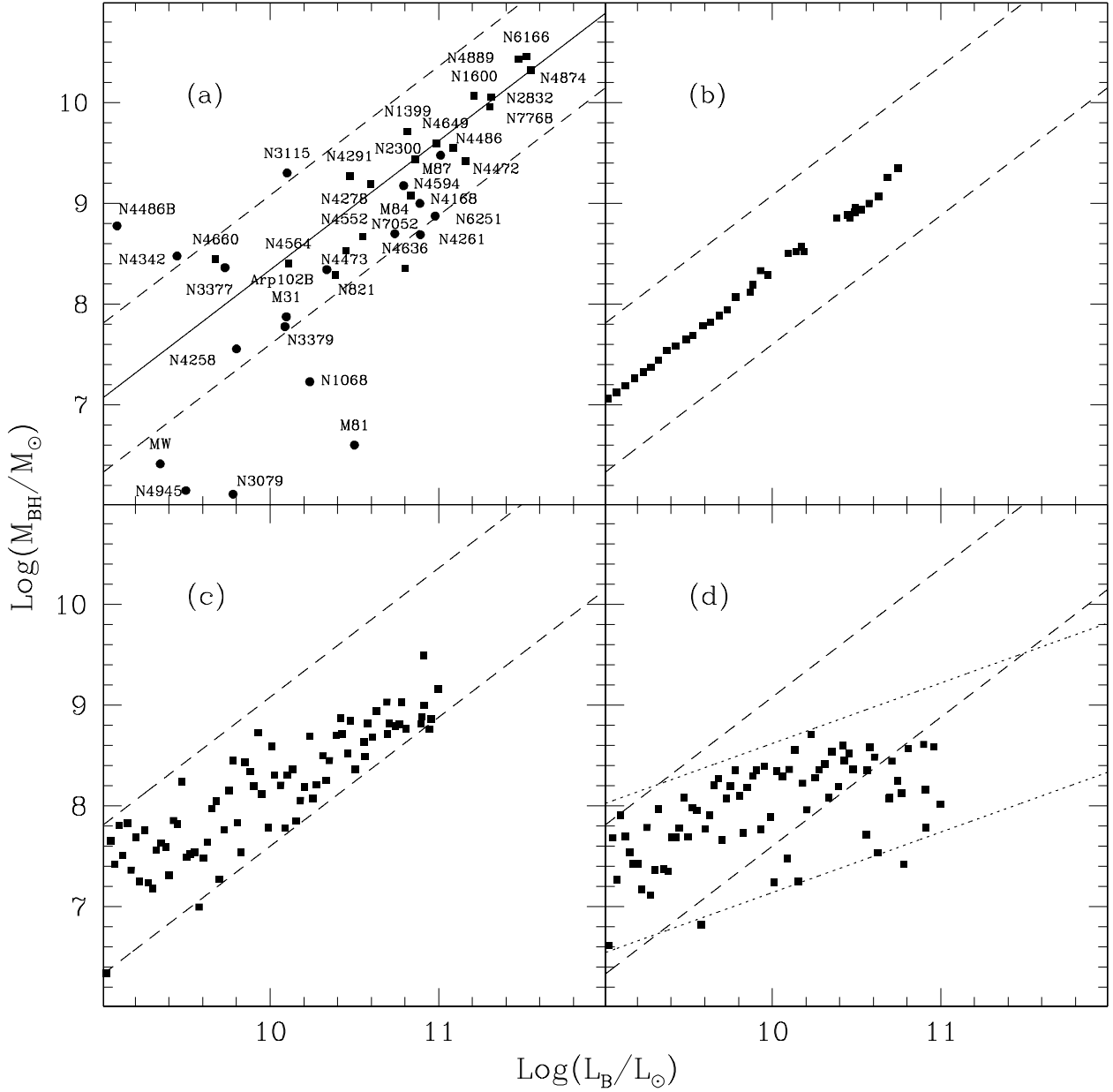,width=\textwidth,angle=0}}
\caption{(a)  Observed black hole masses and bulge luminosities 
for the samples of nearby galaxies compiled by Ho (1998,
{\it circles}) and Magorrian et al. (1998, {\it squares}). The solid line shows the linear least square fit 
to the data, ${\rm log}(M_{BH}/M_\odot)=1.28{\rm
log}(L_B/L_\odot)-4.46$, while the dashed lines show the $\pm 1\sigma$ 
deviation ($\sigma=0.74$). The dashed lines are  the same in all
panels for comparison. 
(b) Monte Carlo simulations of black hole masses and bulge
luminosities with $M_{\rm acc} = 6\times 10^{-3} \Delta M_{\star}$  
     (model i).
(c) Same as (b) with $M_{\rm acc} = 1.4\times 10^{-3} (1+z)^{2}
\Delta M_{\star}$ (model ii).
(d) Same as (b) with $M_{\rm acc} = 10^{-6} (1+z)^{2} 
             M_{\rm halo} \exp{[-(v_{\rm c}/300{\rm\,km\,s}^{-1})^4]}$
(model iii). The linear least square fit to the results of model (iii)
has a slope of 0.6, shown by the two dotted lines.
}
\end{figure*} 

The normalisation  $\epsilon_{\rm acc}$ is determined by requiring  the
Monte Carlo  simulations to reproduce the comoving mass density of 
supermassive black holes at the present time. As discussed in the 
introduction and by HNR98, this value is still uncertain, and
here we have used a value of $10^6M_\odot{\rm Mpc}^{-3}$.
This value is about six times larger than that 
estimated by Choski \& Turner (1992) using optical counts of 
unobscured QSOs.

\section{The mass distribution of supermassive black holes 
in low-redshift galaxies}

Figure 3a shows the observed black hole masses 
and bulge luminosities for a sample of 39 galaxies
as compiled by Ho (1998) and Magorrian (1998),
together with a linear least square fit and the $\pm1\sigma$
deviation. Figures 3b, 3c and 3d  show  the results of 
the Monte Carlo simulations.
Models (i) and (ii) both reproduce the observed roughly linear 
relation between black hole mass and bulge luminosity.
This is mainly due to our assumption 
that the accreted mass is proportional to the mass that 
has formed stars, which is strictly true in model (i) and is still a good 
approximation for model (ii). In model (i) the relation is 
very tight,  tighter than observed. In model (ii), instead, 
the different redshift evolution 
of the mass forming stars and of the mass 
accreted onto supermassive black holes acts as a substantial source of 
scatter, which is now comparable to that in the observed relation. 
Model (iii) produces a shallower relation of
black hole mass to bulge luminosity; 
the predicted remnant black hole masses in large bulges are here  
considerably smaller than observed; this is due to the  
non-linear relation between bulge mass and halo mass: as the assumed 
star formation efficiency depends on the depth of the potential well,
the bulge mass rises faster than the halo mass with increasing bulge
luminosity.
The scatter
in model (iii) is again comparable to the observed scatter and to that
in model (ii).

We have checked the sensitivity of our results to some of our assumptions.
The slope of the relation of black hole mass to bulge luminosity
depends only weakly  on the cosmology, on the details of the
galaxy formation scheme and on the quasar epoch once the parameters 
have been fixed to reproduce the present-day bulge luminosity function 
and the integrated mass density in black holes.
The scatter in this relation,
however, grows when the difference between the redshift evolution
of star formation and black hole accretion increases.  

\section{Discussion and conclusions}

We have used Monte Carlo realizations of the extended Press-Schechter 
formalism together with simple schemes for galaxy formation 
and mass accretion onto supermassive black holes to investigate 
how the  merging and accretion history of supermassive black holes 
within a hierarchical cosmogony affects the mass distribution of
remnant black holes in low-redshift galaxies.  

The observed roughly linear relation between black hole masses 
and bulge masses can be reproduced if the mass accreted 
during a major merger is a fixed fraction of the mass of the gas
that has formed stars in the  merging galaxies since the 
last major merger. This relation is preserved if this 
fraction is independent of mass but varies with redshift, but
such a variation introduces significant scatter into 
the relation. The redshift variation necessary to match the
evolution of the global mass accretion rate inferred from the integrated
light emission of optically bright QSOs (assuming that their 
radiation efficiency does not depend on redshift) results in a scatter 
that is comparable to that in the observed relation.

The linear relation between bulge
and black hole mass and the  modest scatter in this relation 
at the bright end of the bulge luminosity function seem to 
suggest that for most of the integrated mass density in supermassive black 
holes a common mechanism determines the efficiency for mass 
accretion of supermassive black holes and the efficiency for star
formation.

The scatter in the observed relation must be at least partially 
due to observational errors and there seems not much room 
for additional sources of intrinsic scatter at the bright end of the 
luminosity 
function (except in the somewhat artificial model (i)).  
At the faint
end the scatter in the observed relation seems to increase. 
This might indicate that here other parameters, e.g. the spin of 
the DM halo, influence black hole and star formation in a different way. 

If the accreted mass scaled with the mass of the halo rather than with the 
stellar mass, the black hole mass would depend on the bulge mass less steeply
than observed and black holes in luminous bulges would be less massive than
the observations suggest.

\section{Acknowledgements} 
We thank Guinevere Kauffmann and Simon White for comments on the manuscript.
This work was supported by the EC TMR network for ``galaxy formation 
and evolution''. Andrea Cattaneo acknowledges the support of an
Isaac Newton Studentship.

\end{document}